\documentclass[final,5p,times, twocolumn]{elsarticle}

\usepackage{array}
\usepackage{ amssymb }
\usepackage{float}
\usepackage{lineno,hyperref}
\usepackage{url}

\journal{arXiv.org}

\biboptions{authoryear}
\makeatletter
\def\@author#1{\g@addto@macro\elsauthors{\normalsize%
    \def\baselinestretch{1}%
    \upshape\authorsep#1\unskip\textsuperscript{%
      \ifx\@fnmark\@empty\else\unskip\sep\@fnmark\let\sep=,\fi
      \ifx\@corref\@empty\else\unskip\sep\@corref\let\sep=,\fi
      }%
    \def\authorsep{\unskip,\space}%
    \global\let\@fnmark\@empty
    \global\let\@corref\@empty  
    \global\let\sep\@empty}%
    \@eadauthor={#1}
}
\makeatother

\begin{document}

\begin{frontmatter}

\title{A Dynamic Service Description for Mobile Environments}

\author{Rohit Verma\fnref{fn1}\corref{cor1}}
\ead{rohitv@iiti.ac.in}

\author{Abhishek Srivastava\fnref{fn1}}
\ead{asrivastava@iiti.ac.in}

\fntext[fn1]{Computer Science and Engineering, 
		Indian Institute of Technology Indore India}
\cortext[cor1]{Corresponding author}

\begin{abstract}
With the increasing processing capability of mobile platforms and advancements in Internet of Things, modern mobile devices have shown a favorable prospect for on-the-go service provisioning. However, there is much to be done to realize this. A detailed, dynamic, and lightweight service description is an important requirement for automatic and efficient discovery, selection, and subsequent provisioning of services over mobile devices.  Traditional approaches for service description are usually not directly adaptable to mobile environments owing to the latter's dynamic and distinct nature. In this paper, we propose a dynamic, lightweight, extensible, and detailed service description especially designed for mobile environments, considering crucial aspects such as isolated data source, collaborator partners, and hardware aspects along with the functional, non-functional, business, and contextual aspects. The description has been partitioned along these lines and various parts of the description are distributed between service registries and the mobile service providers. An up-to-date and light weight description has been achieved by this, without compromising on the overall consistency of the description. A prototype of the proposed system has been implemented with the intent of validating the feasibility of the approach. Further, the proposed approach is suitable for a heterogeneous environment comprising both wired and wireless systems.
\end{abstract}

\begin{keyword}
Mobile Web Services, Service Description, Service Publishing.
\end{keyword}

\end{frontmatter}

\section{Introduction}

Over the past two decades, mobile technology has gained widespread popularity and has become a part of day-to-day life. 
In particular, smart phones and mobile devices strongly impact the way human beings communicate and deal with digital information.
The modern era has witnessed rapid advancements in the field of mobile technology and wireless networking. As a response to this advancement and growth, a large number of services are emerging in the market that can provide digital information over hand held mobile devices. With such dramatic growth, smart phones and mobile devices have the potential to become
\textit{"service providers"} from merely being \textit{"service consumers"}.

The materialization of the vision to host web services over mobile devices can bring a new level of usability to mobile users. 
The mobile web services will allow the mobile user agent to directly interact with other mobile user agents. This reduction of human intervention in service provisioning will speed up service execution, limit the chances of error, automate redundant tasks, and most importantly reduce the annoyance of human users. 
A few prospective applications of mobile web services are: 
1) Credit cards, debit cards, visiting cards can be provided as web services from mobile devices without the need of having the user search for them or even carry them physically. 
2) Localization of personal information can be done seamlessly through services over mobile devices.
3) Modern mobile devices are equipped with powerful sensors. Mobile devices laden with such sensors play the role of a ``gateway'' facilitating proper access to the capabilities of the sensors.
4) Mobile services are particularly useful in scenarios where there is little or no preexisting infrastructure by functioning through ad-hoc networks. Examples of these scenarios are war-front, post-disaster relief.

The realization of web services over mobile devices has gained attention in the community. Several works have been proposed to provide web services over modern mobile devices~\citep{mobile2002}\citep{mobile2003}\citep{mobile2006}\citep{mobile2010}\citep{mobile2011}.
However, a key challenge that is overlooked here is~\textit{``service description''}.
Service description is crucial for the consumers of services to get a sense and better understanding of the offered services and operations. 
This is of further importance in mobile environments, where the service invocation requires a great deal of service understanding owing to the dynamic nature of transient services.
Traditionally, WSDL (Web-Service Description Language) is used to describe and publish the technical description of web-services. Well written WSDL documents provide binding implementation information, detailed description of input-output messages, information on how messages are sent through the network, amongst others. 
Further, WSDL documents facilitate discovery of the intended web-services over standard service registries such as UDDI - Universal Description Discovery and Integration. 
On the flip side, however, WSDL documents only provide the functional information of a web-service. 
They do not provide information on the \textit{non-functional aspects}, \textit{contextual aspects}, and the \textit{business aspects} of a web-service. This information is crucial and of utmost importance in selection and proper usage of available services especially in the context of providing such services over mobile devices. 

In an earlier work of ours~\citep{Rohit2015SCC}, we were able to incorporate functional, non-functional, contextual, and business aspects of services to service descriptions for mobile devices. In this work, we plan to build upon that work to incorporate several more important aspects to such descriptions of mobile services. These include descriptions of service collaborators, data source details, hardware aspects, and consumer base. 
Mobile devices may sometimes act as the ``gateway'' to information provided by data sources such as embedded sensors, third party applications, or other mobile services. 
In such scenarios, the mobile service provider and the data source can be viewed as two separate entities. 
Both the entities are operated autonomously, with their own unique characteristics and further both entities are prone to failure independently. 
In such scenarios, traditional service description solutions do not suffice as they consider data source and service provider as indistinguishable entities. 
An important perspective covered by the proposed mobile service description is: ``data source'' and ``mobile service provider'' are looked upon as two disjoint and independent entities.
To the best of our knowledge, this paper is the first attempt at considering the service provider and data source as two separate and autonomous entities and proposes a service description solution accordingly.


We extend our earlier approach to incorporate descriptions on data source and decouple the descriptions of these two entities (service provider and data source). Further, we acknowledge the fact that mobile web services are usually light weight and provide limited functionality. 
Mobile services can be combined and aggregated among themselves to build and compose more complex and useful services.
Hence, this collaboration can offer to provide services that can be readily useful in a real world scenario.
In the proposed approach, we further incorporate details on the collaborative partners. 
The goal of this paper is to provide various service descriptions and information handy to the service consumer.
Access to such information along with technical descriptions at the time of service discovery speeds up the service selection process considerably. Further, this can facilitate the service consumers to shortlist and select the most suitable and optimum service provider beforehand without the need to communicate with individual service providers.
The trade-off though in making such additional information available is that it proportionately increases the size of the description document. The mobile service consumers need to perform heavier processing to handle such lengthy descriptions. Furthermore, such description information (which could include availability, location, response time, latency, price, service scope) is subject to frequent changes owing to the very nature of mobile environments. Management of such detailed service description documents at the service registry therefore could easily suffer from consistency issues, lack of up-to-date information, and increased network traffic. In this paper, we attempt to tackle such issues and provide a feasible solution for mobile service descriptions.

%
%
%
%
%
%
%
%

The aim is to provide a lightweight solution that is dynamically update-able and facilitates rich service descriptions in mobile environments. We emphasize, however, that the proposed solution is not a \textit{replacement} for existing technologies but one that complements it. It acknowledges the heterogeneity of the environment that supports a  co-existence of wired, wireless, and mobile devices. The idea is to extend the WSDL 2.0~\citep{wsdl20}, to incorporate non-functional, contextual, business, data source, collaborator information. The extension takes into account the constraints and issues of  mobile environments. To the best of our knowledge, this is the first attempt at providing a lightweight yet exhaustive service description solution that facilitates dynamic updates in mobile environments.

The rest of the paper is organized as follows: a summary of the concepts used for service description in mobile environments is presented in Section~\ref{sec:background}. A motivating scenario for mobile SOA and service description is presented in Section~\ref{sec:scen}. A brief description of the problem forms Section~\ref{sec:problem}. Detailed discussion on the extensible, dynamic service description is included in Section~\ref{sec:sol}. Evaluation of the proposed approach is presented in Section~\ref{sec:eval}. Finally the related work section and concluding remarks form parts of Section~\ref{sec:related} and Section~\ref{sec:conclude}.

\section{Background}
\label{sec:background}


In this paper, we use the term mobile services to imply self-contained and reusable services that are provided by \textit{mobile devices} or sometimes by \textit{human users} via mobile devices.
Such mobile services are application components that facilitate device to device communication over mobile environments.
They provide means to communicate between various software applications of mobile devices.
Such services may be utilized for commercial and non-commercial purposes. 
A few applications of mobile services are:

- Replacement of physical day-to-day things: Physical credit cards, debit cards, identification cards, access keys can be provisioned as mobile services hosted over mobile devices of users.

- Gateway to sensor provided information: Mobile devices can provision services that offer information provided by general purpose sensors (e.g. Location sensor) or special purpose sensors (e.g. Medical sensors (ECG sensors, Body glucose sensors)), Environmental sensors (Fire sensors, Barometer sensor). 

- Personal information provider: Services that offer information of a person can act as a dynamic and digital visiting card. This can help him to socialize without the need of introducing himself again and again. Further, this can be used by applications that record attendees information in a particular meeting.

- Service in Infrastructure-less environment: Mobile services are particularly useful in the scenarios where there is little or no preexisting infrastructure by functioning through mobile ad-hoc networks. Examples of these scenarios are war-front, post-disaster relief.

Though such services are already provided over wired networks and are widely used, the concept of services hosted and provided over mobile devices is relatively new and is in an inception phase. Services offered over mobile devices of users has the potential to substantially reduce the total cost of ownership for hosting a service. Further such mobile services are able to provide personalized and contextual services/information in a more effective manner. 
One may also suggest running a web service on somewhere in the wired infrastructure and making the mobile device a proxy for the same rather than actually hosting it over the mobile device. 
Though running mobile services as a proxy can work in some situations, there are many applications and scenarios (as discussed above) that require services to be present on the mobile device.
Further, firewalls sometimes block the access between mobile services and proxy. Also, provisioning, configuring, and maintaining a proxy is harder for a common user than maintaining a device itself. Hence, hosting web services over mobile devices is a better approach in most scenarios.~\cite{mobile2003} provides an interesting discussion on the same.

\begin{figure}
\centering
\includegraphics[scale=0.45]{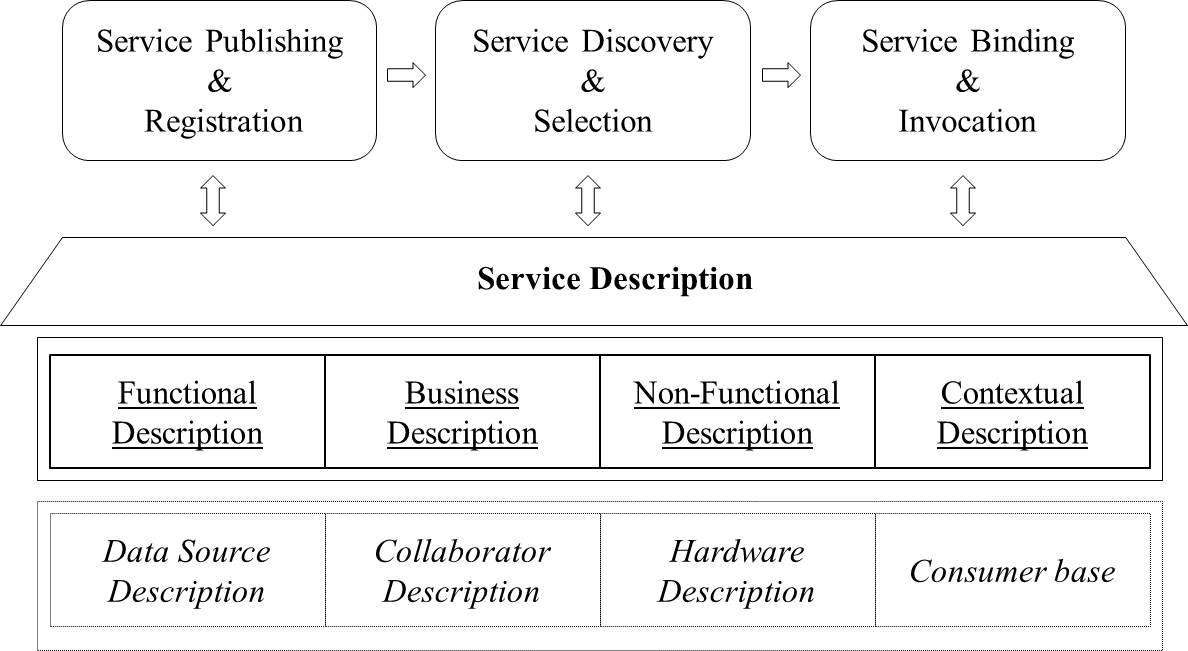} 
\caption{Role of Service Description}
\label{fig:problem}
\end{figure}

An effective way to make the most of mobile services and easing service consumers interaction with mobile services is a well defined ``Service Description''. 
This enables service consumers to effectively discover and use the offered mobile service. 
Figure~\ref{fig:problem} depicts the role of service description in service systems. 
In a service system, the usual stages are: Service Publishing, Service Discovery, Service Selection, Service Binding, and Service Invocation. 
Service description is an integral part in most of these stages and therefore the role of service descriptions can not be overemphasized.
%
A service description is a means to express the characteristics of the offered service to unknown prospective consumers. 
Well expressed descriptions are important for service consumers to make sense of the offered services and operations. 
Service descriptions provide clear and structured instructions on how to invoke a service which is particularly important to first time service consumers. 
Moreover, an exhaustive service description eases device-to-device communication by automating various stages of service systems (as shown in Figure~\ref{fig:problem}). 
Hence, in this paper our prime focus is a service description framework for mobile environments.



In the proposed work, we propose a solution to complement the existing technology particularly WSDL. 
We work towards extending the features of the WSDL document to accommodate the needs of the mobile environment. 
WSDL already provides a concrete fundamental technology that can be adapted for mobile environments
to describe and publish functional and technical parts of the mobile services.  
In the current context where wired legacy systems and modern hand-held mobile devices coexist, a service description framework that works well in the heterogeneous mixture of diverse technologies and provides a platform for interoperability is required.

To the best of our knowledge WSDL 2.0 is best suited for such requirements. WSDL 2.0 is capable of describing both the major web service technologies: \textit{SOAP based and REST (Representational State Transfer) based}; this is possible as WSDL 2.0 has good support for describing HTTP bindings. WSDL 2.0 further provides a generic mechanism to define service operations using Message Exchange Patterns (MEP)~\citep{wsdl20}. This feature encourages message-oriented operations and supports arbitrary message exchanges that are pertinent for heterogeneous mobile environments. Although several other description languages have been proposed since the inception of WS-* technology, most do not take into account the distinct nature of mobile environments. This, along with the wide acceptance of WSDL has made us rely on it for the functional description of services in mobile environments. Furthermore, continuing with WSDL would require least tinkering with existing protocols and technologies. 

Being XML based, WSDL 2.0 has very convenient in-built  extension capabilities that can sufficiently cater to our requirements.
Our approach is to extend WSDL 2.0 to accommodate various other description aspects in addition to the functional description that it already takes care of. For this, we have utilized the ``import'' statement of WSDL 2.0 for linking physically separate description documents. These partitioned descriptions enable lightweight, dynamic, and consistent management of the overall service description.

\section{Motivating Scenario}
\label{sec:scen}

In this section, we present a brief scenario that demonstrates the importance of the non-functional, contextual, and business aspects of service description in addition to the usual functional description. 
Further the scenario discusses about the importance of the data source, collaborator, and hardware description in a mobile environment:

\textit{Alice is going on a solo budget trip to Paris for the first time. It was a last minute decision and therefore her trip is not well planed. As she lands at Paris Airport, her mobile device fetches for her the available accommodation options. 
It discovers a few lodging options from the function descriptions of the mobile web services exposed by the mobile devices of Bob (hostel owner), Carol (solo traveller), Dave (group traveller), and Eve (Tourist Agent). 
However, her mobile device is unable to shortlist the best option from these for her. This is because it is not aware of various other vital details such as service reliability, throughput, network capacity (Non-functional Description), current online status, location, battery status (Contextual Description), price to access, security, minimum version for client agent (Business Description). Nonetheless, upon Alice's intervention the accommodation services offered by Bob are selected. As she moves towards Bob's accommodation, her mobile device fetches for her the nearby food options. Apart from the caf\'{e}, restaurants, her mobile device also discovers a few innovative food buddie services offered by other travellers (Carol, Frank, and Ron) hosted over their mobile devices that seek a partner for an existing dinner reservation. Her mobile device, however, was unable to shortlist the food buddie services further as the details on the restaurants they are interested in, its menu, reservation policy, and other descriptions were not available (Collaborator Description). Therefore, her mobile provides Alice a vanilla list of offered services.
Alice likes the food buddie service as she wants to socialize on her nomadic trip.
Therefore, she decides to join Carol, but later Alice regrets her decision as she had to travel to the other part of the new city. On her way back, her mobile device discovers a few mobile services offering information on a shorter route with less traffic. However, the source of the routing information is unknown (Data Source Description). Therefore, the services were not used despite being genuine. 
Alice's mobile device discovers a few mobile services offering discounted entry tickets for the Eiffel Tower and the Louvre Museum. However, her mobile device is not able to further shortlist these services because apart from the functional description no any other description was available. Hence, it requires some manual intervention that introduces human biases and chances of accessing multiple services for getting things done.}

Now, if the updated non-functional, contextual, and business information were made available to Alice along with the functional service description, it would have saved Alice the hassle of communicating and negotiating with irrelevant service providers. These aspects of description are therefore worth considering in a mobile scenario: non-functional description would give Alice an idea about the overall performance of a mobile service; business related constraints or information (such as availability of service in her locality, usage price for service, service background i.e. human provided, in-house developed service, sensor provided service) would be provided to Alice through the business description; contextual description would brief her about the context of the service offered. 

Furthermore, these mobile services may offer services on the basis of data provided by an external entity (e.g. in case of shortest route services GPS sensors and map services are used). In such scenarios, information of those data sources becomes crucial. Moreover, information about the service provider's hardware is also important to assess the feasibility of the service and claimed service QoS from the provider's hardware. 
This information (non-functional, contextual, business, data source, hardware, collaborator), however, cannot be archived in the service registry along with the functional description. This is because mobile devices are by nature inherently mobile and hence these descriptions continuously vary. There needs to be, therefore, a mechanism to support and ease the frequent updates of the service description.

In the following sections, we summarize the problem and demonstrate our solution that offers wider (covering functional, non-functional, contextual, business, data source, collaborator, and hardware aspect), light-weighted and update-able service descriptions to handle the challenges in mobile environments.

\section{Problem Statement}
\label{sec:problem}

Several heterogeneous \textit{mobile devices} constitute the mobile environment. These devices could be of varying processing capabilities, power requirements, memory, transmission protocols.
Further, these devices are prone to uncertain behavior and dynamic changes as they usually are in continuous motion and they can randomly join/leave the network. 
Therefore, a holistic service description mechanism is required for services hosted by such mobile devices that comprehensively covers the various unique aspects of the mobile environment. 
Merely a functional service description does not 
provide enough information to the service consumer for service selection in such environments.
Service selection made on the basis of only functional service description may lead to the invocation of an obsolete service or an off-line service provider.
 
A novel approach is required that takes into account the distinct nature of mobile environments and that considers service aspects in addition to the functional description, such as the non-functional, contextual, data source description, hardware, and business descriptions of a mobile service. 
In the current scenario \textit{legacy wired systems} and \textit{modern wireless mobile systems} coexist. 
Therefore, a completely novel architecture that intends to replace the existing solutions is undesirable. 
A service description solution that complements the current solutions and also has added features catering to mobile environments is the need. 
As elaborated in the earlier sections, in mobile environments the service descriptions require frequent updates owing to regular context changes, non-functional and/or business related changes. 
Moreover, mobile environments are prone to failures reasons being frequent network outages, battery limitations and are usually constrained in terms of processing power. This leads to the added requirement of a `lightweight approach' for such service descriptions. 
Further, keeping in mind the distinct nature of mobile environments, information about the data source, collaborator, and hardware becomes an important parameter during service selection.
Hence, a service description approach that considers the issues associated with mobile environments and at the same time blends with the existing technological solutions is required. 

We summarise the requirements of mobile service description as: detailed (including non-functional, contextual,  business, collaborator, data source, hardware aspects in addition to functional); run-time update-able (i.e. dynamic); and lightweight.

\section{Proposed Approach}
\label{sec:sol}

\begin{figure}
\centering
\includegraphics[scale=0.35]{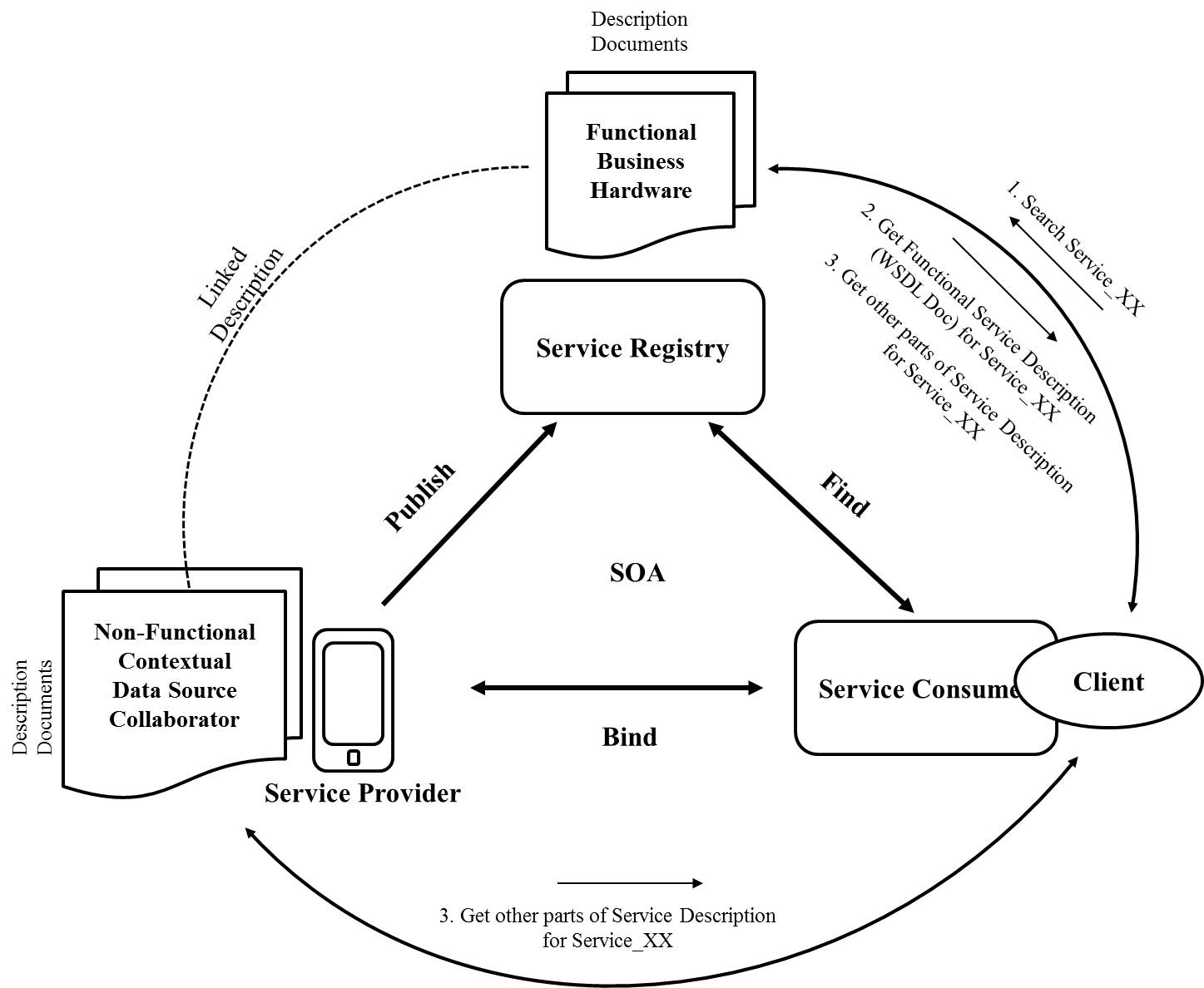} 
\caption{Mobile Service Description}
\label{fig:alterdescp}
\end{figure}

As discussed, we intend to use and extend the existing technologies and tools for service description in mobile environments.
We propose to extend WSDL for service description to accommodate the 
distinguished requirements and features of services in mobile environments. 
In the subsequent subsections, we provide a brief overview of the proposed approach:

\subsection{Design Concept:}
\label{subsec:concept}

We propose to use the following service description documents for mobile services: 

\begin{enumerate}
\item Functional Description Document.
\item Non-functional Description Document.
\item Contextual Description Document.
\item Business Description Document.
\item Data Source Description Document.
\item Collaborator Description Document.
\item Hardware Description Document.
\end{enumerate}


WSDL documents are widely used for service descriptions. These documents 
provide detailed functional or technical description of services. Hence, we rely on WSDL documents for functional description of the services provided over mobile devices.
In this work, we propose to link the WSDL document with other description documents mentioned above using the ``import'' statement defined in WSDL. 
The ``import'' mechanism allows referral to other WSDL documents defined elsewhere.
We use this to connect descriptions that are split across multiple documents for the mobile service. 

Figure~\ref{fig:alterdescp} shows an abstract view of the proposed approach. There are three primitive entities: Service Provider (mobile device hosting a service), Service Consumer (mobile or non-mobile device), and Service Registry (mobile registry or traditional non-mobile registry). These entities have the Publish/Find/Bind relationship between them as shown in Figure~\ref{fig:alterdescp}.
The service descriptions are split into multiple documents and placed at the Service Registry and the Service Provider (i.e. the mobile device in this case).
The motive behind this splitting and placing of multiple parts of the description at different locations is: 

\begin{itemize}
\item to facilitate faster, independent, and dynamic updates related to service provider in the descriptions and keeping the description up-to-date. 
\item to maintain the overall consistency of the description in case of simultaneous updates.
\item to provide a lightweight and detailed service description. 
\end{itemize}

The descriptions of a mobile service may be dynamic and subject to updates regularly. This includes descriptions related to the current network (Wi-Fi or GSM), location, current availability status of the mobile service provider hosting the service. 
These updates are managed by various independent entities or authors and, therefore, the mutual independence of separate description documents saves the hassle of inconsistent updates. 
For instance, the non-functional description of a mobile service can be managed and updated by a third party auditor or a broker, whereas
the contextual description can be managed by a simple mobile application residing on the same device.
This demonstrates the efficacy of splitting and delocalising service description documents in the mobile environment.

Registration of services hosted over mobile devices with a service registry is assumed to follow the same process as in traditional systems. In case of UDDI, the service registration involves the bindingTemplate, businessEntity, businessService, publisherAssertion, tModels~\citep{alonso2004web}. Our approach extends and makes use of WSDL that primarily affects the tModels. We are not proposing any change to the fundamental structure of the WSDL. Hence, the existing process for service registration and publishing is adapted.
The proposed approach, therefore, blends well with wired systems and existing SOA standards.
The main differences, however, lie in the service description retrieval process.
Following are the steps used in description retrieval in the proposed approach (as shown in the Figure~\ref{fig:alterdescp}):

\begin{enumerate}
\item Search a service at the service registry.
\item Retrieve the functional service description from the service registry.
\item Retrieve the rest of the service descriptions from the mobile service provider.
\end{enumerate}


\begin{figure*}
\centering
\includegraphics[scale=0.40]{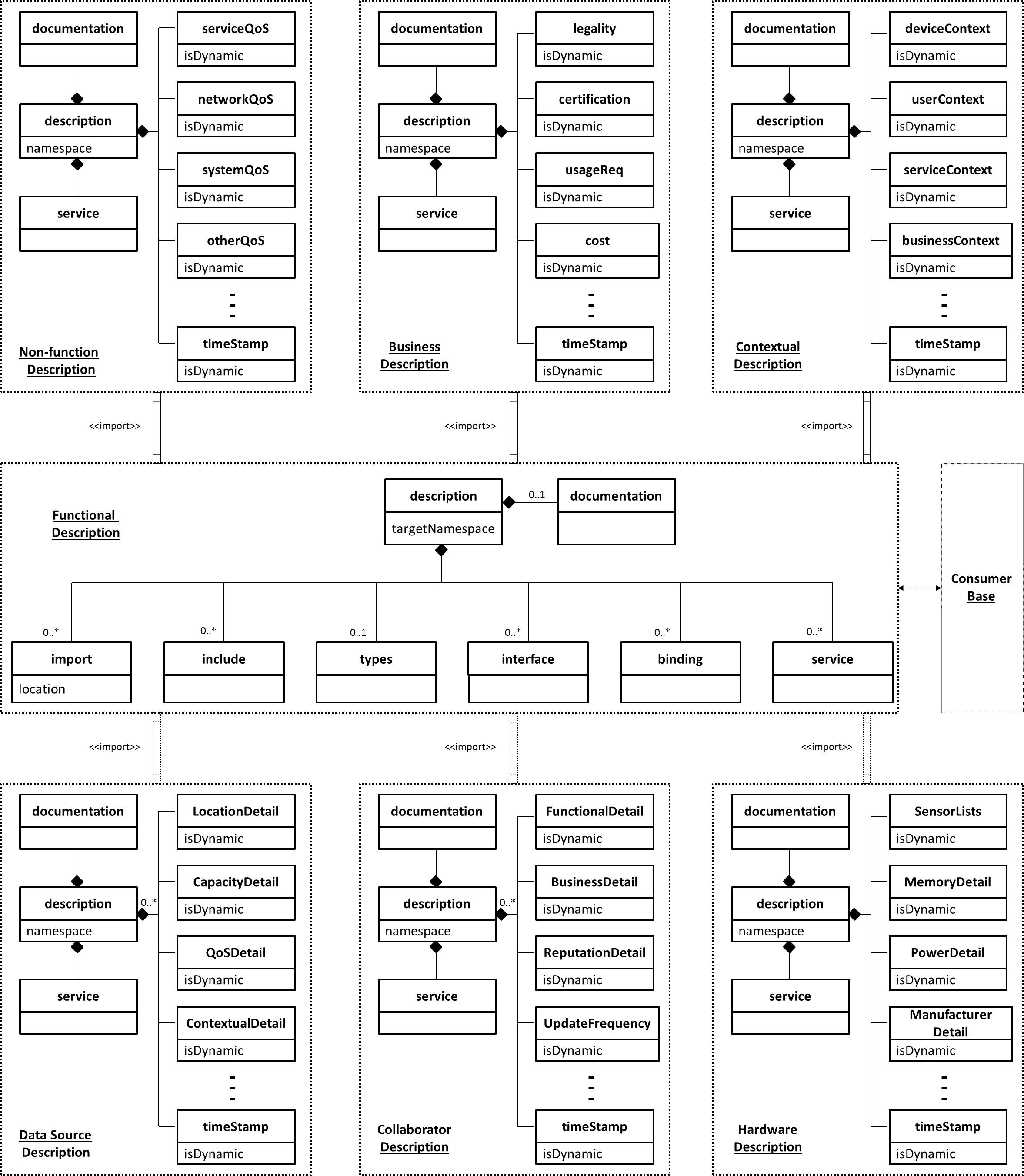} 
\caption{Service Description Infoset for Mobile Services}
\label{fig:exten}
\end{figure*}

\subsection{Description Components:}
\label{subsec:component}

As discussed in Subsection~\ref{subsec:concept}, we propose to use various service description documents that provide a holistic understanding of a mobile service and that describe the 
mobile provisioned services in a distributed manner.
As shown in Figure~\ref{fig:exten}, seven service descriptions are interlinked by the \textit{import} statement and an information base about the service consumers. 
Each element of the descriptions has an attribute \textit{``isDynamic''}, that indicates whether the element is dynamic or not i.e. if the element requires regular updates. 
This particularly helps in cases where a third party broker or application is responsible for updating the description. 
In this paper, we include several description metamodels for mobile services, however, defining all the elements associated with each description metamodel is out of the scope of this paper.

A brief overview of each service description is as follows:

\subsubsection{Functional Description}
\label{subsubsec:funct}

As discussed in the earlier subsection, we rely on WSDL 2.0 for a detailed functional description of the service (refer to Figure~\ref{fig:exten}). A functional description should describe ``What'', ``How'', and ``Where'' of a service:

\begin{itemize}
\item \textit{What} - The function of a service or the operations that a service provides is described by the functional description of a service. The \textit{interface} component of the functional description (refer to Figure~\ref{fig:exten}) helps to achieve this.
\item \textit{How} - The mode of invoking the service, the message formats, and transmission protocols used by the service constitute an important part of the function description. The \textit{binding} component of the functional description (refer to Figure~\ref{fig:exten}) helps to achieve this.
\item \textit{Where} - The location of a service or the URI of the service endpoint conveys to the service consumer the address of the service. The \textit{service} component of the functional description (refer to Figure~\ref{fig:exten}) and the endpoint element in it defines the URI of a service deployment.
\end{itemize}

This description is usually the first criterion during the service selection process. 
First of all, the services are shortlisted on the basis of the functionality they provide, subsequent filtering of the services is done through further criteria.
In the proposed approach, therefore, we link the remaining descriptions of the service with the functional description using the ``import'' statement. A typical import statement comprises  a namespace and the location of the importing description document: 
\textless \begin{verb} import \end{verb} \begin{verb} namespace="anyURI" \end{verb} \begin{verb} location="anyURI" \end{verb}\textgreater \textless \begin{verb} documentation /\textgreater \textless \end{verb}\begin{verb} /import \textgreater \end{verb}

%

\subsubsection{Non-Functional Description}
\label{subsubsec:nonfunc}

The non-functional properties or the quality of service implies the overall performance of the service experienced by the service consumers. 
We propose to associate a ``timeStamp'' attribute with the non-functional description document that indicates the time-stamp of the last update. 
In the uncertain mobile environment, the non-functional properties change with a change of context of service (e.g. location, network, battery etc.). Hence, the time-stamp brings in a degree of certainty to the non-functional description. 
This helps in better management of dynamically varying description elements. We propose to use the ``timeStamp'' attribute with the  business and contextual descriptions as well.

We propose four `Quality of Service' (QoS) groups in mobile environments: 

\begin{enumerate}
\item \textit{serviceQoS:} Service QoS are the quality attributes of a service as experienced by service consumers. A few important QoS attributes include: Availability, Capacity, Latency, Throughput, Performance, Reliability.

\item \textit{networkQoS:} Network QoS include the quality attributes associated with the underlying network used by the service. This network varies from Wireless LAN, GSM, WiMAX. A few attributes of this group are: 
Packet Loss, Network Delay, Delay Variation, Bandwidth Capability.

\item \textit{systemQoS:} System QoS are the quality attributes that characterize the whole system instead of just the service or network or third party application.
A few attributes in this category are: Accessibility, Security, Usability, Scalability, Interoperability, Robustness (Failure-Management), Extensibility. 

\item \textit{otherQoS:} This group or placeholder is proposed to categorize the quality attributes that do not fall in any of the groups mentioned above. This group is extensible and can further categorize service attributes. A few examples of this group are: Testability, Modifiability, audit-ability.

\end{enumerate}

This description is not only limited to these four types and can be extended to include other QoS attribute categories as well. A few pointers to work on non-functional properties of services are~\citep{QualityAttributes2005}\citep{o2006towards}\citep{extWSDL2006}\citep{Kritikos2009}. We consider the QoS as the totality of the features and characteristics of the service that are based on its ability to satisfy the implied needs (as per ISO 9000).
In this paper, we only focus on the description of the claimed or known QoS values of the offered services. 
Determination/Estimation of the varying QoS attribute values is beyond the scope of this work.

\subsubsection{Business Description}
\label{subsubsec:business}

The business related information of a service is expressed in the business description document. This description mainly comprises:
\begin{itemize}
\item \textit{Legality}: The legal obligations or conditions associated with the service are represented by this placeholder.
For instance, a service is not available in a specific country, the service is making use of proprietary applications, disclaimer notifications. The legality description is particularly important in case of mobile services as they can easily migrate from one legal boundary to another.
\item \textit{Certification}: The certification placeholder specifies the business related certifications or licenses associated with the service. For example ISO certification, SSL security certificates.
\item \textit{Usage Requirement}: The preconditions for service usage (if any) and other service usage requirements are described by the usage requirement placeholder. This may include the minimum version of software agents or device capabilities. 
\item \textit{Cost}: The Cost or pricing placeholder specifies the price for use of the service. The cost placeholder could further be extended to cover discounts, special offers, group pricing for a set of services.
\end{itemize}

Apart from this, the business description document could include information related to referred service choreography, service offering background (e.g. 
in-house development, human provided service, mobile sensor provided service or third party applications), service version\citep{tempoWSDL2012}, service scope (what a service covers and what it does not). 
The business description information is necessary in mobile environments as it provides greater exposure to the business related offerings of mobile services.

\subsubsection{Contextual Description}
\label{subsubsec:context}

The often varying context of mobile devices makes the contextual information of utmost importance for services provisioned over mobile devices. A good definition of the term context is given by Bazire and Br\'{e}zillon~\citep{context2005}:
``Context is any information that can be used to characterize the
situation of an entity, where the entity is a person, place, or object that is
considered relevant to the interaction between a user and its application, including
the user and the application themselves''.

The context information includes constraints, nature, and structure of the factors that influence the behavior of the service.
The description chiefly has placeholders for the following elements:
\begin{enumerate}
\item \textit{deviceContext:} This represents the context of the mobile device that hosts and provisions the service. This includes device related information such as sensors, battery status, device data plan. 
\item \textit{userContext:} This placeholder depicts the user (mobile device owner) or (human service provider) related information. User context is worth considering in
mobile environments as the mobile device user's (or human service provider's) activities, behavior can directly influence the service experienced by the service consumers. This may include the user's routine, availability of the service, the user's background (e.g. profession), user's situation (walking, running, driving), location (address, GPS coordinates, time zone), presence.
\item \textit{serviceContext:} Service related contextual information is depicted by this placeholder. A few examples of service context are service domain, service connection preference, service specialisations.
\item \textit{businessContext:} Business contextual information includes information such as 
preferred business scenario (e.g. combination of user's and device's context), preferred service partners, compositions.
\end{enumerate} 

\textit{Service} and \textit{Documentation} are common attributes in all descriptions. Service specifies the associated service name and its URI for which the description has been provided, while documentation specifies the human readable 
descriptions of the attributes and service description. These two attributes are borrowed from WSDL 2.0.



\subsubsection{Data Source Description}
\label{subsubsec:InfoSource}

In mobile services, mobile service providers (or mobile devices) often serve as the ``gateway'' to information provided by the data source which is itself an independent entity. 
The data source can be anything that provides the mobile service with data.
The data source and mobile service provider can be viewed as two independent entities that have independent failures, context, capacity. 
Examples of such data sources are internal sensors that are physically located within mobile devices (GPS sensor, digital compass, barometer, pedometer) or external sensors (smart home sensors, body area network sensors etc.) or third party mobile software applications. 

As technology progresses towards the Internet of Things (IoT), there is expected to be rapid increase in the number of such data sources. The mobile service provider will then more commonly provide an abstraction for data obtained from such `things'. The abstraction would be such that the data is made available in formats that are standard and facilitates seamless usage.
While for the service consumers, the description of these data sources would provide a better understanding and a holistic view of the service. That, consequently, might become an important service selection criteria. The data source description primarily contains placeholder for the following elements:

\begin{enumerate}
\item \textit{LocationDetail:} This comprises the location details of the data source. This includes the GPS coordinates and other the location information.
\item \textit{CapacityDetail:} This comprises the technical details on the data source. This mainly includes the 
physical capacity of the data source. In certain scenarios, this may include battery information and computation capacity as well.
\item \textit{QoSDetail:} This provides non-functional information on the data source. This may include availability, throughput, reliability, network delay, security information.
\item \textit{ContextualDetail:} Contextual information (as discussed in the earlier point) provides information about the constraints, nature, and structure of the factors that influence the behavior of the data source.
\end{enumerate}

The data source descriptions are provided and managed by the mobile service provider. This document could further be split into two parts: Dynamic Description (Placed in the vicinity of the service provider) and Static Description (Placed at the service registry). This description is necessary as it provides greater exposure to the important constituents of the mobile service. 


\subsubsection{Collaborator Description}
\label{subsubsec:CollabPartner}

Mobile devices are powerful enough to provide services on their own, yet their capabilities can be improved manifold through mobile service collaboration. Description and information on collaborators helps prospective service consumer to take a decision on a service provider.

Collaborator description provides information and conformity to the service consumer about the service being offered.
In this paper, we provide the following placeholders for collaborator description:

\begin{enumerate}
\item \textit{FunctionalDetail:} This placeholder is proposed to provide information on what, how, and where of the service collaborator. The functional description of the collaborator (as discussed in section~\ref{subsubsec:funct}) may be reused. 
\item \textit{BusinessDetail:} This placeholder is proposed to provide business related information on the collaborator. This includes legality, usage requirements, cost, and other related aspects. 
\item \textit{ReputationDetail:} This placeholder provides information on the reputation of a collaborator. This is mainly based on feedback and updates from the previous service consumers and other certified reputation assessment sources (\cite{wang2007}).
\item \textit{UpdateFrequency:} This placeholder specifically works towards prevention of out dated information in a large workflow. This enables the service consumer to have an updated service all the time. 
\end{enumerate}

\subsubsection{Hardware Description}
\label{subsubsec:Hadware}

Hardware details of the service provider are provided in this placeholder. 
This description is introduced to minimize the need of service negotiation from non-potential providers. 
Service consumers can assess the service claim and the hardware that is used to provide the service before actually using the service. The following placeholders are introduced for a detailed hardware description:

\begin{enumerate}
\item \textit{SensorLists:} Modern mobile devices are equipped with several modern sensors. This placeholder provides a detailed list of the equipped sensors and their functionality.
\item \textit{MemoryDetail:} This briefs the service consumer about the memory details of the mobile device providing the service. Memory details include information on the primary memory and secondary memory of the mobile device. In certain scenarios, this placeholder also includes details about external memory locations (in case cloud storage is used).
\item \textit{PowerDetail:} Power or battery plays an important role in the selection of mobile services. This placeholder 
provides the runtime power profile of the mobile device.
\item \textit{ManufacturerDetail:} This placeholder provides information on the manufacturer, kernel versions, and other device related information. This could further provide information on the manufacturer of the mobile processor, WiFi and bluetooth adapters. 
\end{enumerate}

Hardware related details may be fetched at runtime using APIs exposed by the modern mobile operating systems. For example android provides detailed and sophisticated libraries that access the hardware information efficiently. This description document could further be split into two parts: Static Part (comprises the unchanging elements - SensorList, ManufacturerDetail and is placed at the service registry) and Dynamic Part (comprises the changing elements - MemoryDetails, PowerDetail and is placed in the vicinity of the service provider).

\subsubsection{Consumer Base Details}
\label{subsubsec:ConsumBase}

The consumer base provides details about the earlier consumers of the service. This helps prospective service consumers better assess a service provider. We can further extend this placeholder to include feedback and rank the service providers. These can be used to propose a recommendation system for mobile services. Detailed discussion on such recommendation systems is beyond the scope of this paper.

\section{Evaluation}
\label{sec:eval}

We have evaluated the proposed approach with the rationale of demonstrating its usability, feasibility in practical scenarios, and efficacy for mobile service description. We have used the following evaluation techniques (as discussed in~\citep{Keng2011}): 
1) Feature Comparison, 2) Empirical Evaluation, and 3) Theoretical and Conceptual Evaluation.
Section~\ref{subsec:comp} provides a detailed feature comparison, Section~\ref{subsec:experiment} discusses the empirical evaluation, Section~\ref{subsec:case} makes use of case studies for theoretical evaluation.

\subsection{Feature Comparison}
\label{subsec:comp}

In this section, we provide a thorough comparison of the proposed approach with existing service description methods available in literature. For this comparison, we examine 22 related approaches and compare the proposed approach on the following criteria: applicable domain, ability to dynamically update the description, representation style, aspect of description covered, techniques used for description, validation approach for the method, and other salient features. Table~\ref{tab:comp1} and Table~\ref{tab:comp2} present a detailed comparison of the proposed approach and existing works chronologically in literature .

\begin{table*}
\centering
\caption{Comparison of Proposed Approach with Existing Approaches in Literature}
\label{tab:comp1}
\begin{minipage}{.90\linewidth}
\scalebox{0.8}
{
\begin{tabular}{|m{5.4cm}|m{2.2cm}|m{1.2cm}|m{2cm}|m{2.5cm}|m{1.5cm}|m{3cm}|}
\hline 
\textbf{Work} & \textbf{Domain} & \textbf{Dynamic Update} & \textbf{Representation} & \textbf{Technique Used} & \textbf{Validation Approach} & \textbf{Description Aspect} \\
\hline
Web Service Description Language~\citep{wsdl2001} & Wired & No & Syntactic & XML & W3C Specification & Functional \\ 
\hline
DAML-S~\citep{Ankolekar2002} & Wired & No & Semantic & Darpa Agent Markup Language(DAML+ OIL) ontology & Example & Functional \\
\hline
WSDL extension for Security description of Web services~\citep{wsdlexten2002} & Wired & No & Syntactic & XML & Not Mentioned & Functional \\
\hline
Security Description Framework~\citep{Morioka2003} & Wired, Wireless & Yes & Syntactic & XML & Example & Functional and Security Description\\
\hline
OWL-based framework of the Semantic Web~\citep{owls2004} & Wired & No & Semantic & RDF & W3C Specification & Functional\\
\hline
Formal Service Description Language~\citep{Hartmann2006} & Wired & No & Syntactic and Semantic & ForSeL Calculus & Case Study & Functional \\
\hline 
Situation Aware Service based Systems~\citep{yau2006} & Wired & No & Syntactic & Extention of OWL-S with situation ontology (SAW-OWL-S) & Example & Functional and Contextual \\
\hline
Model driven WSDL extension~\citep{extWSDL2006} & Wired & No & Syntactic & XML & Example & Functional and Non-Functional \\
\hline
Semantics for service description~\citep{pfeffer2007} & Wired & No & Semantic & Distributed Semantic Tree (DST) & Not Mentioned & Functional \\
\hline
WSDL-Lite~\citep{vitvar2007} & Wired & No & Syntactic and Semantic & RDF Schema & Example & Functional, Non-Function and Behavioral\\
\hline
SOAP Service Description Language (SSDL)~\citep{Fornasier2007} & Wired & No & Semantic & SSDL & Not Mentioned & Functional\\
\hline
WSMO-Lite~\citep{Vitvar2008} & Wired & No & Semantic & SAWSDL annotations & Not Mentioned & Functional \\
\hline
Web Application Description Language~\citep{wadl2009} & Wired, Wireless & No & Syntactic & XML & W3C Specification & Functional \\
\hline
WSDL extension for version support~\citep{Juric2009} & Wired & No & Syntactic & XML & Prototype & Functional and Technical \\
\hline
Unified Service Description Language~\citep{USDL2010} & Wired & No & Syntactic & MOF-based meta-model & W3C Specification  & Functional \\
\hline
WSDL extension for non-functional attributes~\citep{extWSDL2010} & Wired & No & Syntactic & XML & Case Study & Functional and Non-Functional \\
\hline
Intentional approach to service description~\citep{Rolland2010} & Wired & No & Intentional (From business Perspective) & Intentional Service Modeling for Service Description & Example & Functional\\
\hline
WSDL extension for criteria support~\citep{extWSDL2011} & Wired & No & Syntactic & XML & Case Study & Functional \\
\hline
WSDL-temporal~\citep{tempoWSDL2012} & Wired & No & Syntactic & XML & Case Study & Functional \\
\hline
Service description with extended semantic and commercial attributes~\citep{keppeler2014} & Wired & No & Semantic & XML & Prototype & Functional and Commercial \\
\hline
Context Aware Mobile Cloud Services~\citep{OSullivan2015} & Wired, Wireless & No & Syntactic & JSON and HTTP & Prototype & Functional \\
\hline
WSDL extension for mobile environment~\citep{Rohit2015SCC} & Wired, Wireless & Yes & Syntactic & XML & Prototype & Functional, Business, Non-Functional, Contextual \\
\hline
WSDL extension for holistic mobile service description (Presented Approach)& Wired, Wireless & Yes & Syntactic & XML & Prototype, Case Study & Functional, Business, Non-Functional, Contextual, Data Source, Collaborator, Hardware \\
\hline
\end{tabular}
}
\end{minipage}
\end{table*}

\begin{table*}
\centering
\caption{Salient Features of Existing Approaches in Literature}
\label{tab:comp2}
\begin{minipage}{.90\linewidth}
\scalebox{0.85}
{
\begin{tabular}{|m{5cm}|>{\centering\arraybackslash}m{14cm}|}
\hline 
\textbf{Work} & \textbf{Salient Features}\\
\hline
\cite{wsdl2001} & Web Service Description Language is a W3C recommended XML based interface definition language. WSDL provides machine readable description of the functionality that are offered by a web service. The current version is WSDL 2.0.\\ 
\hline
\cite{Ankolekar2002} & This work provides semantics to the web service by the use of DAML+OIL ontology. The objective being making web services computer interpret-able and enabling service discovery, invocation, inter-operation, composition, verification, and other operations semantically enriched. \\
\hline
\cite{wsdlexten2002} & This work extends WSDL and UDDI to incorporate security features. The extension facilitate both public-key and trust policy. The description supports publication of various security parameters such as provider encryption, public key signatures, access control policies, and data usage policy. \\
\hline
\cite{Morioka2003} & The paper presents a scalable security description framework for mobile web services. WSDL is extended such that change in the service context changes the channel security level and AAA service flow. \\
\hline
\cite{owls2004} & OWL-S provides semantic web description for web services and enables semantic based automated service discovery, invocation, and composition. OWL-S was formerly known as DAML-S. \\
\hline 
\cite{Hartmann2006} & The work presents a formal service description to represent functional aspects of a service. The services are considered to be ``re''-action system that is activated when input is triggered and some precondition holds.\\
\hline
\cite{yau2006} & This work presents a situation based service aware system. An extension to OWL-S has been presented for situation ontology that is incorporated in service specification.\\
\hline
\cite{extWSDL2006} & The WSDL extension is proposed to incorporate QoS characteristics of a web service in the description. Model driven architecture recommendations were used to carry out meta-model transformation. \\
\hline
\cite{pfeffer2007} & This work presents semantic service description. A light-weight semantic service description is proposed using distributed semantics trees. These trees are hierarchical representation for service effect descriptions. \\
\hline
\cite{vitvar2007} & The work presents an extended web service description stack that adds a semantic layer to the service description. Web service modeling language is used to express the service description semantics. \\
\hline
\cite{Fornasier2007} & The work presents an alternative web service description language SSDL. SSDL provides a lightweight solution for SOAP based web services, it is a message-centric approach that fits with SOA systems. \\
\hline
\cite{Vitvar2008} & A minimal lightweight ontology for semantic web services is presented. SAWSDL was used to define WSMO-lite for arbitrary semantic description. \\
\hline
\cite{wadl2009} & Web Application Description Language is designed to provide machine readable, XML based descriptions to the HTTP based web applications and RESTful services. \\
\hline
\cite{Juric2009} & This work presents extension of WSDL to support versioning of web service. Service-level and operation-level versioning is handled in the proposed work. \\
\hline
\cite{USDL2010} & Universal Service Description Language is a proposed for describing business, operational, and technical aspects of the universal services. It is a general purpose, domain independent language for Internet of Services. \\
\hline
\cite{extWSDL2010} & A flexible extension of WSDL is proposed to incorporate non-functional attributes of a web service. Model driven architecture is used to extend the WSDL meta-model. Further few of the requirements of IOT have also been addressed.\\
\hline
\cite{Rolland2010} & The work presents intentional level service description. Intentional service model is presented to describe the intentional services and register them with the service registry. \\
\hline
\cite{extWSDL2011} & WSDL is extended to X-WSDL that included `criteria' as the non functional property of a web service. `criteriadefinition' and `criteriaservice' keyword has been included in the extended WSDL.\\
\hline
\cite{tempoWSDL2012} & WSDL-temporal is proposed as an extension of WSDL to manage the issues related to the change management in the web services. The proposed method allows to multiple version of the interfaces within a single web service. \\
\hline
\cite{keppeler2014} & A model to include multiple attributes for semantic and commercial information in service description is proposed. A holistic XML-based description language along with primitive prototype is discussed. \\
\hline
\cite{OSullivan2015} & A user oriented service description is proposed to allow simple user interaction without any technical details. Provision for both REST and SOAP is discussed. The primary focus of the work is to use cloud services from mobile phones using a in-build cloud assistant.  \\
\hline
\cite{Rohit2015SCC} & This work proposes a light weight and extensible approach for service description, that is designed for mobile environments. Dynamic update to the service description is proposed to keep the description up-to-date.\\
\hline
Presented Approach & WSDL 2.0 is extended to provide description for mobile hosted services. Further, descriptions are distributed to the service registry and service providers on the basis of their update frequency. Functional, Business, Non Functional, Contextual, Data Source, Collaborator, Hardware descriptions are proposed to provide a holistic description for mobile services\\
\hline
\end{tabular}
}
\end{minipage}
\end{table*} 

%

\subsection{Empirical Evaluation: Prototype}
\label{subsec:experiment}

We put together a working prototype for assessing the feasibility of the proposed approach.
Actual mobile devices were used to deploy the prototype.
The prototype is capable of managing dynamic service descriptions in a mobile environment. 
For this, an android application was developed that is capable of communicating with service registries, retrieving the description documents, extracting relevant information from these descriptions, and updating the descriptions dynamically. 
The Android operating system was chosen to implement the prototype because it is open source, has wide market share and availability.
The proposed approach is generic and can be extended for use on any mobile platform. Our experimental setup comprised four mobile devices (including two Samsung Galaxy S Duos with Android 4.0, Google Nexus 7 with android 5.0, and Asus Zenfone 5 with Android 4.4), one laptop (Intel i3 2.13 GHz with 3GB of RAM) and a few instances of the prototype running on virtual instances of android devices on the laptop.

\begin{table}[h]
\centering
\caption{Proposed Features and its Prototype Realization}
\label{tab:1}
\begin{minipage}{\linewidth}
\scalebox{0.9}
{
\begin{tabular}{|>{\raggedright\arraybackslash}m{1.55cm}|>{\raggedright\arraybackslash}m{6.5cm}|}
\hline
\textbf{Proposed Features} & \textbf{Realization in Prototype}\\
\hline
Lightweight & Service descriptions are pulled from the service registry and the service providers on requirement. XML parsing is done by the android toolkit's in-built DOM parser. Used android ``service''$^1$ majorly for development.\\
\hline
Runtime Update & Several Android API's$^2$ are available to get the battery, network, location, sensor details. A watchdog process is created to update the description with the latest information.\\
\hline
Detailed Description & Managed various descriptions on various documents located at the service provider. Descriptions placed at the service provider provide detailed updated business, contextual, and claimed non-functional information (Figure~\ref{fig:exten}).\\
\hline
Extensible & Considered description meta-models/infoset in the description documents, with an option to introduce new parameters in these meta-models based on service description requirements. \\
\hline
\end{tabular}
}
\end{minipage}
$^1$http://developer.android.com/guide/components/services.html
$^2$http://developer.android.com/reference/android/package-summary.html
\end{table}

\begin{figure}
\centering
\includegraphics[scale=0.3]{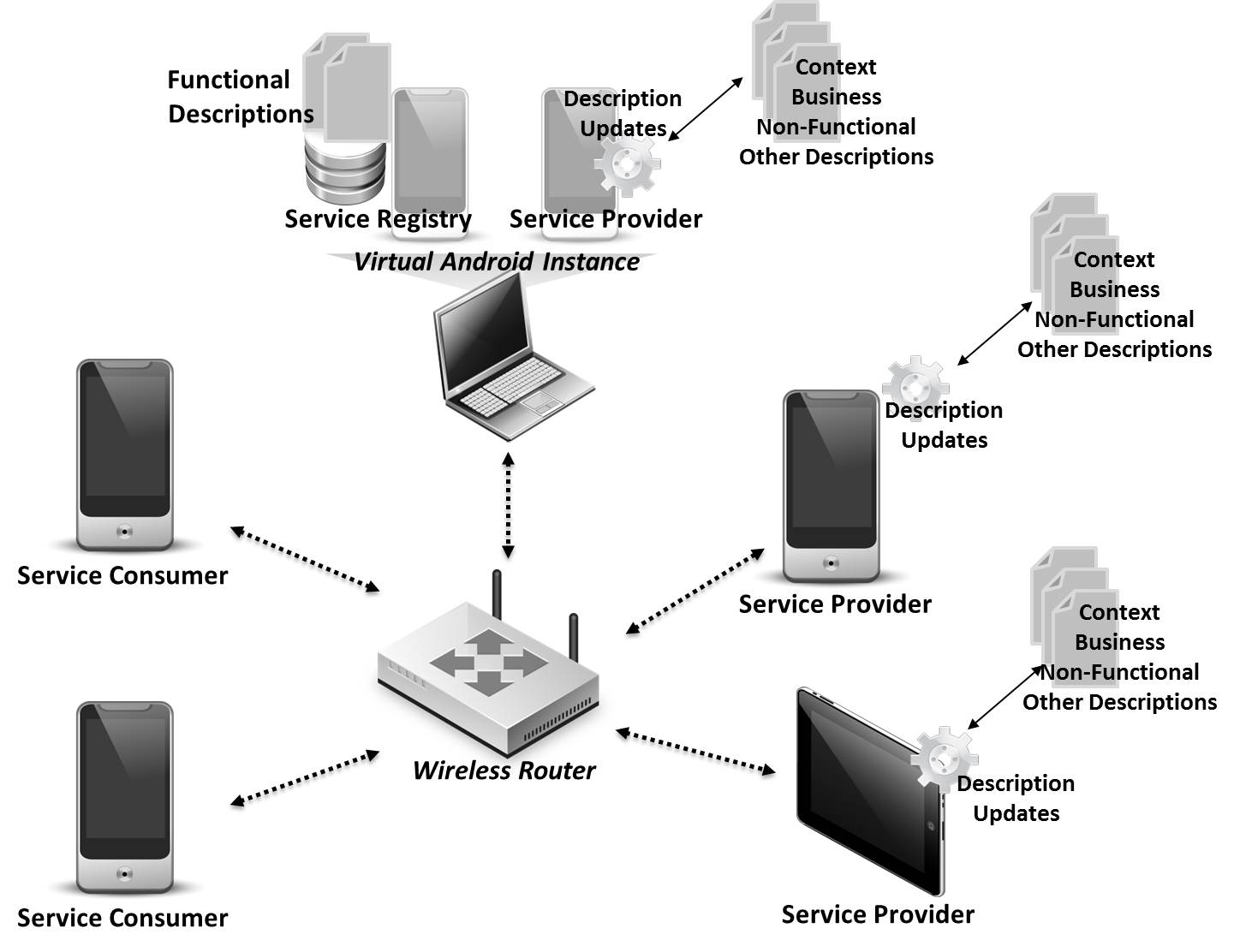} 
\caption{Mobile Service Description Prototype}
\label{fig:proto}
\end{figure}

\begin{figure}
  \centering
  \includegraphics[scale=0.2]{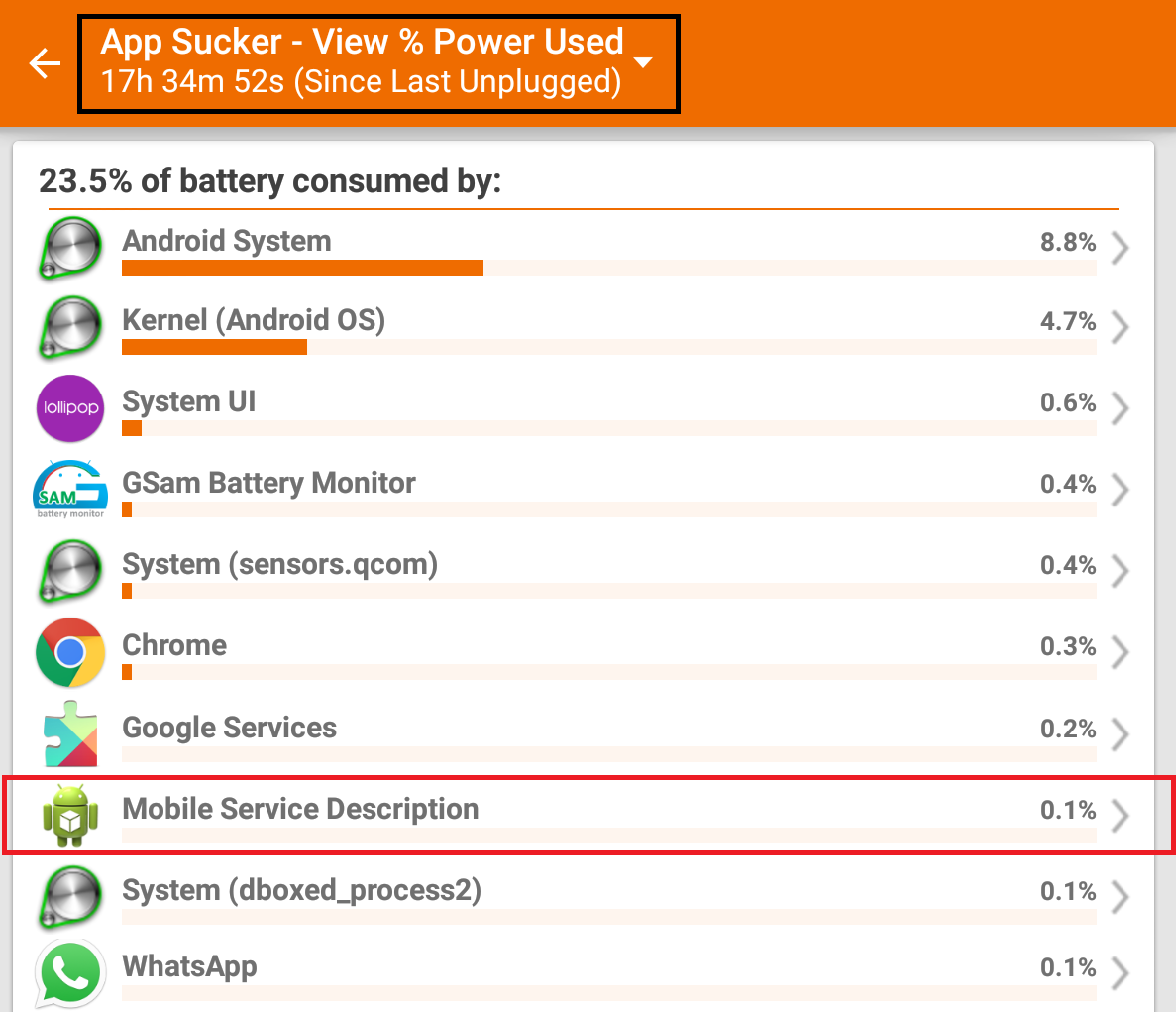}
  \caption{Prototype Battery Usage on the Android Device}
  \label{fig:sub2}
\end{figure}

The proposed framework was evaluated in a practical setting. 
We requested four volunteers to deploy the prototype over mobile devices and roam around within the institute campus. 
We established an experimental wireless network within the institute's building to connect the volunteers' mobile devices, laptop, and its running virtual instances. This is shown in Figure~\ref{fig:proto}.
During the experiment, the mobile devices followed a random pattern of mobility as the volunteers did not follow any predefined roaming pattern. The battery usage by the prototype on one of the android devices is shown Figure~\ref{fig:sub2}. This shows minimal battery usage and our prototype also had a small memory footprint of 1.14MB. This clearly indicates the feasibility of the prototype for use in the real world. A brief overview of how the various features of the proposed approach were realized in our prototype implementation is given in Table~\ref{tab:1}.

Two mobile devices and a virtual android instance played the part of the service provider and hosted services along with the description documents as discussed earlier. Further, we engineered a `watchdog' application to sense the changes in the service provider's contextual, business, and non-functional information and accordingly kept the description documents updated. We made use of a mobile based service registry~\citep{rohit2014} and hosted the functional description documents over it. The mobile devices acting as service consumers retrieved the functional description document (i.e. WSDL document) from the service registry. Subsequently, the consumers extracted the location information on the other description documents (viz. business, contextual, non-function) from the same WSDL and these were retrieved.

We also analysed the UDDI registry to identify if any tweaking was required to accommodate the requirements of the proposed approach. For this, we deployed Apache jUDDI version 3.1.5 on CentOS Linux server 6.2 and SoapUI version 4.6.3 on a Windows 7 machine. One of the concerns was related to the fact that we were using WSDL 2.0 and some of the existing UDDI registries are meant for WSDL 1.1. However, this turned out to be a non-issue as most  WSDL 2.0 documents can be mapped to WSDL 1.1 documents. We conclude that given the fact that we made use of WSDL as the base description and that UDDI already maps the WSDL through tModel (as discussed in Subsection~\ref{subsec:concept}), there is no requirement for major change to the UDDI registry in the move to accommodate the proposed framework.

\begin{figure}
  \centering
  \includegraphics[scale=0.2]{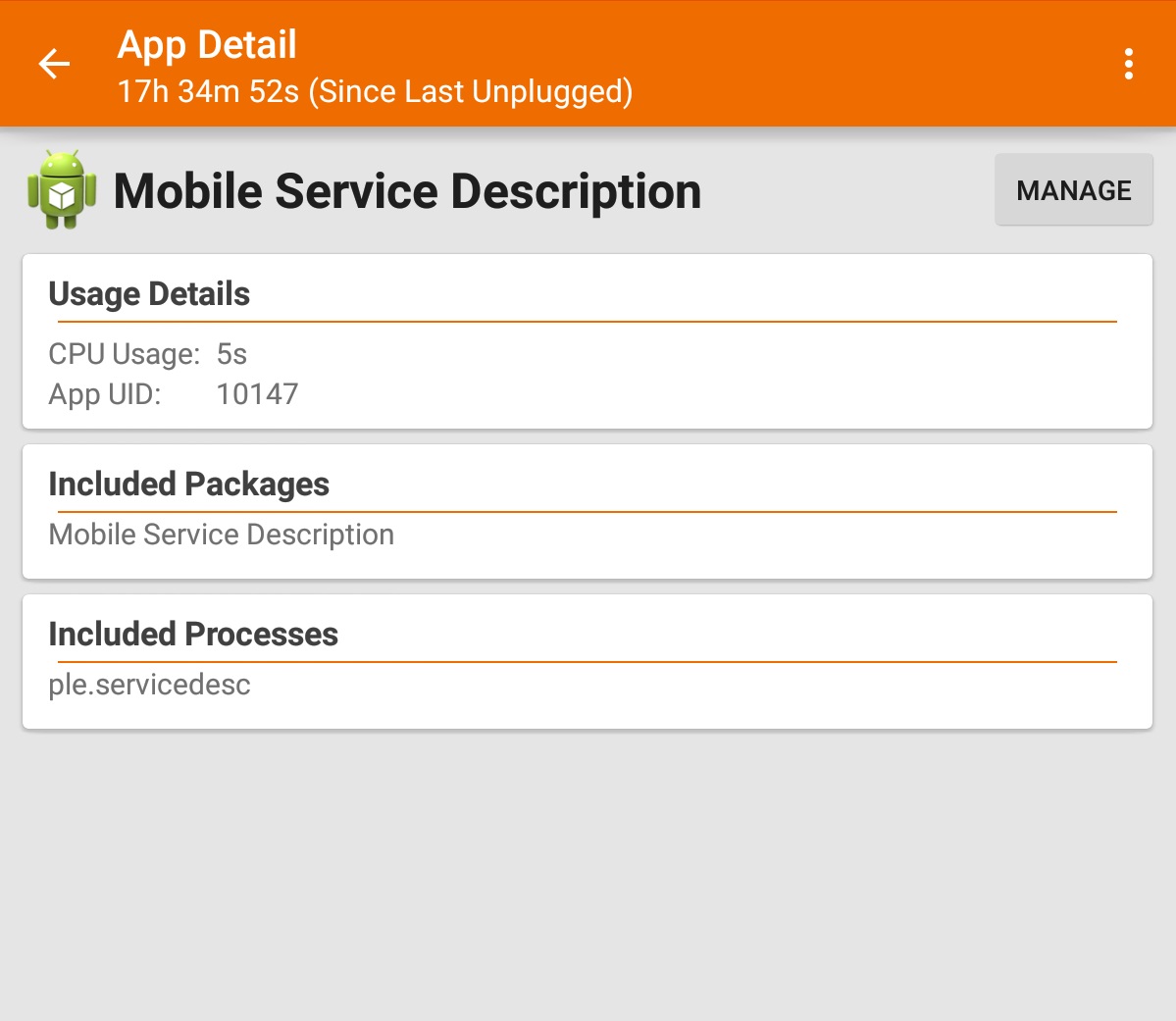}
  \caption{Prototype CPU Usage on the Android Device}
  \label{fig:sub3}
\end{figure}

\subsection{Conceptual Evaluation: Case Study}
\label{subsec:case}

In Section~\ref{subsec:experiment}, we described a real world deployment of the proposed approach by developing a working prototype and deploying it over real mobile devices. In order to further validate the proposed service description, we conduct a case study of the same in this section.

We acknowledge that the conventional wisdom about case study research has several prejudices. These prejudices have been discussed in detail in an interesting article "Five Misunderstandings About Case-Study Research" by Bent Flyvbjerg~\citep{flyvbjerg2006}. Going along the points mentioned in this article, we present 
three different service case studies that may exist in the service ecosystem. First, we discuss the case of a shopping mall where the services provide information on the latest offerings: \textbf{MallLatestOffer}. Second, we discuss the case of a \textbf{SalesmanTracking} mobile service. And third, we take the example of a \textbf{CarPoolingMate} service. Table~\ref{tab:case1} discusses these cases briefly. The motive behind discussing these three examples is to assess our description approach for three types of primitive mobile services: 
1) Automated Mobile Services: Services that are offered by mobile device itself and do not involve the human. Example- Mobile services that offer sensor provided information, Mobile services that offer personal information viz digital visiting card.
2) Semi-Automated Mobile Services: Services that are provisioned over mobile devices that sometimes requires human intervention. Example- Mobile service that offers meeting availability for a person along with its GPS location. 
3) Manual Mobile Services: Services that are offered by human and mobile devices act as a gateway or interface for their services. Example- Mobile service interface for human provided services~\citep{Schall2012}.

\begin{table}[!htbp]
\centering
\caption{Mobile Services Case Study Details}
\label{tab:case1}
\begin{minipage}{.90\linewidth}
\scalebox{0.87}
{
\begin{tabular}{m{2.2cm}m{6cm}}
\hline
\textbf{Service Name} & \textbf{Service Details}\\
\hline
MallLatestOffer & \textbf{Type}: Semi-Automated Mobile Service \\
 & \textbf{Dependencies}: Other services from Mall\\
 & \textbf{Functions}: Provides latest offers from various brands of the Mall. Make use of existing services of brands that provides offer details and provides the offer information manually if offer service is not available.\\

 & \\
SalesmanTracking & \textbf{Type}: Automated Mobile Service \\
 & \textbf{Dependencies}: GPS sensor, Mapping Service \\
 & \textbf{Functions}: Provides location information of the salesman that helps the manager to track the salesman's location and plan their next visit. This make use of mobile phone's GPS sensor and mapping service.\\

 & \\
CarPoolingMate & \textbf{Type}: Manual Mobile Service \\
 & \textbf{Dependencies}: None\\
 & \textbf{Functions}: Provides the carpooling information. This helps the traveler to fetch the car pooling mate may be in a meeting or a remote public function. This requires to provide the information manually at the provider's end. \\

\hline
\end{tabular}
}
\end{minipage}
\end{table}

We perform requirement coverage analysis of the proposed description approach for these case studies in Table~\ref{tab:case2}. This coverage analysis helps us assess whether the proposed mobile description is required for various types of mobile services and whether the proposed description meets the unique description requirements of various classes of mobile services i.e. Automated mobile services that make use of the device's sensor or other services, semi-automated services that may use other services and sensors and also make use of manual information/data supply, and manual services that requires manual supply of information/data (human-automation continuum). 

Based on our analysis with these case studies, all services definitely require a functional description. Most other descriptions discussed in the paper are usually also required by most services. The exceptions to this are description about data-source and the collaborator which are not necessary for manual mobile services as there is no other collaborator involved.




\begin{table}
\centering
\caption{Mobile Service Description Requirement Coverage for Case Studies}
\label{tab:case2}
\begin{minipage}{\linewidth}
\scalebox{0.75}
{
\begin{tabular}{llccc}
\hline
 & & & \textbf{Case Study}  & \\[0.1cm] \cline{3-5}  
\textbf{Service Description} & & MLO$^1$ & ST$^2$ & CPM$^3$ \\[0.1cm]
\hline

Functional Description & Include & \checkmark & \checkmark & \checkmark \\
 & Types & \checkmark & \checkmark & \checkmark \\
 & Interface & \checkmark & \checkmark & \checkmark \\
 & Binding & \checkmark & \checkmark & \checkmark \\
 & Service  & \checkmark & \checkmark & \checkmark \\
 
Non-functional Description & ServiceQoS  & \checkmark & \checkmark & \checkmark \\
 & NetworkQoS  & \checkmark & \checkmark & \checkmark \\
 & SystemQoS  & \checkmark & \checkmark & \checkmark \\
 & OtherQoS & \checkmark & \checkmark & x \\

Business Description & Legality & x & \checkmark & \checkmark \\
 & Certification & \checkmark & \checkmark & x \\
 & UsageRequirement & \checkmark & \checkmark & \checkmark \\
 & Cost/Price & \checkmark & \checkmark & \checkmark \\

Contextual Description & DeviceContext & \checkmark & \checkmark & \checkmark \\
 & UserContext & \checkmark & \checkmark & \checkmark \\
 & ServiceContext & \checkmark & \checkmark & \checkmark \\
 & BusinessContext & \checkmark & \checkmark & \checkmark \\

Data Source Description & LocationDetail & \checkmark & \checkmark & x \\
 & CapacityDetail & \checkmark & \checkmark & x \\
 & QoSDetail & \checkmark & \checkmark & x \\
 & ContextualDetail & \checkmark & \checkmark & x \\

Collaborator Description & FunctionalDetail & \checkmark & \checkmark & x \\
 & BusinessDetail & \checkmark & \checkmark & x \\
 & ReputationDetail & \checkmark & \checkmark & x \\
 & UpdateFrequency & \checkmark & \checkmark & x \\

Hardware Description & SensorList & x & \checkmark & x \\
 & MemoryDetail & \checkmark & \checkmark & \checkmark \\
 & PowerDetail & \checkmark & \checkmark & \checkmark \\
 & ManufacturerDetail & \checkmark & \checkmark & \checkmark \\

\hline
\end{tabular}
}
\end{minipage}
$^1$MLO - MallLatestOffer 
$^2$ST - SalesmanTracking 
$^3$CPL - CarPoolingMate 
\end{table}

\section{Related Work}
\label{sec:related}

Service description is an important mean that provides service specification to the prospective service consumers. 
Although literature emphasize the necessity of service descriptions for mobile web services, the unique service description requirements for mobile services is often overlooked. Existing literature rely on the traditional approaches of service description for mobile services, however, these approaches fall short to cater the specific needs of the mobile environment.


One of the most prominent and widely used description language is WSDL~\citep{wsdl2001}.
It has been used traditionally to describe  wired web services.
WSDL describes various technical perspective of web services including service, interface, operations, endpoint, binding, and type definition. Despite the fact WSDL being effective and popular, it does not cover various other aspects of the service specifications (non-functional, contextual, business, data-source) that are pertinent to the mobile environments.

As WSDL is capable of providing functional and technical information, some of the works focused on extending WSDL to incorporate unavailable properties while relying on WSDL to describe technical aspect. 
One of the earlier work from Adams and Boeyen\citep{wsdlexten2002} extended WSDL that introduced security of web services in the description itself. They added optional security parameters to the WSDL and UDDI in order to provide a secure web service transaction.
D'Ambrogio~\citep{extWSDL2006} introduced QoS characteristics of web-services in description by proposing lightweight extension of WSDL. 
In this inspirational work, author made use of metamodel transformation and model driven architecture (MDA).
Versioning of web service interfaces was introduced in description by Juric et al.\citep{Juric2009}, WSDL extension was proposed to support versioning of service interfaces at development-time and run-time.
Agarwal and Jalote~\citep{exten2009} proposed extensions to WSDL and suggested end-to-end support for non-functional properties description, measurement, and update.
Parimala and Saini\citep{extWSDL2011} proposed an extended WSDL and specified the criteria as non-functional properties of web-services.
Change management was focused in the work of Banati et al.\citep{tempoWSDL2012}. WSDL extension (WSDL-Temporal) was proposed to handle issues related to change management. Their approach suggested management of multiple accessible versions of a web-service.
Some other work~\citep{extWSDL2010, extWSDL2011}  also suggests incorporation of non-functional attributes to the WSDL as well. Amongst recent work \citep{extWSDL2014} extend WSDL  for describing complex geodata in GIS services.

O'Sullivan \citep{o2006towards} presents a domain independent taxonomy for conventional services and web services, that is capable of describing non-functional properties. There work provides the ability to communicate non-functional properties along with the service descriptions.
Scheithauer et al.\citep{Scheithauer2009} presents various perspectives and service properties to specify service description.
Zachman framework was used for specifying service properties and their relationship from a service provider's viewpoint for service descriptions depending on the relative perspective.
Cardoso et al.\citep{USDL2010b} and Charfi et al.\citep{USDL2010} proposed a new service description language named Unified Service Description Language (USDL). USDL supports human and IT supported services and provides a domain independent description. However, we feel that an entirely new technology is constrained owing to lack of support for legacy systems. 

Recently there has been several approaches proposed for cloud services. Gal\'{a}n et al.\citep{Galan2009} proposed a service specification language based on OVF (Open Virtualization Format) standard for cloud computing platforms. Sun et al.\citep{Sun2011} presents a description for cloud resources of cloud service provider, thus enabling the cross-cloud implementations. Liu and Zic \citep{Liu2011} proposed cloud\# to provide the service delivery transparency and enhance the trust of cloud service users. This cloud service specification focused on describing how services are delivered inside a cloud. Sun et al.\citep{survey2012} presents an interesting survey of service description languages from the point of view of cloud computing.

A few other related endeavours include: an ontology related to the context explained in~\citep{context2004}, Dustdar's survey on context aware web-service systems~\citep{contextSurvey2009},~\citep{poveda2010context} context ontology for mobile environments.~\citep{mobContext2007} is an inspiring work by Dorn and Dustdar that suggests three types of context for mobile web-services: User-related Context, Service-related Context, Task-related Context.
\citep{knutson2005publishing} is a patent that makes use of WSDL and proposes multi-parted description. Although it only 
describes functional description in multiple documents. 


As stated earlier, a detailed discussion and comparison of the proposed approach and existing methods is already presented in 
Table~\ref{tab:comp1} and Table~\ref{tab:comp2} of this paper. We have studied the existing approaches for service description from the point of view of mobile devices.

Though most of these works do not provide generic specification to fit functional, non-functional, contextual and business aspects of services.
Moreover, these works do not cover tackle the specific requirements of dynamic update, separate specifications for data sources, and collaborators.
To the best of our knowledge, no existing work  is especially dedicated to incorporate the intricacies of the mobile environment. Our work is the first attempt to propose a service description mechanism for such environments.





%


\section{Conclusion}
\label{sec:conclude}

In this paper, we present a novel, lightweight, dynamic, and extensible mechanism for service description especially designed for services hosted over mobile devices. The proposed service description facilitates automated service discovery, selection, and composition. The approach is designed around WSDL 2.0 with the intent of making it useful across both wired and wireless environments.

The mobile environment is very dynamic and it is normal for service description attributes to change frequently over time. 
We proposed the partitioning of the mobile service description into multiple parts: Functional Description, Non-functional Description, Business Description, and Contextual Description.
Further, we added descriptions to facilitate better assessment of mobile services by service consumers such as data source information, collaborator description, hardware details.
The parts of the description that tend to change regularly are made local to the mobile device hosting the service. The motive is to enable seamless dynamic updates in service descriptions without compromising on the overall consistency of the description. 
The proposed solution has the potential to further ease the service selection for prospective service consumers. Further, the solution has the potential to help consumers confine service shortlisting and would avoid the obsolete and irrelevant services. 

Future work in this direction would be towards investigation of additional properties for mobile services and the application of the proposed approach for service selection in the field of service oriented crowd sourcing. We also plan to design a dynamic querying model for more efficient description search in a distributed mobile environment. We further plan to extend our approach to include semantic descriptions for mobile services. 



\section*{Acknowledgment}

The authors would like to thank fellow colleagues, to help
with the technical issues faced during experimentation, and figuring out with the technicalities
of service description. We would like to especially thank the anonymous reviewers of SCC 2015 for their valuable feedback and comments on our initial work. This work is supported by the Ministry of Human Resource Development - Government of India.

\section*{References}

\bibliography{Mdesc}

\end{document}